\documentclass[preprint,showpacs,aps,nofootinbib]{revtex4}

\usepackage{graphicx}% Include figure files
\usepackage{dcolumn}% Align table columns on decimal point
\usepackage{bm}% bold math
\usepackage{latexsym}
\usepackage{epsfig}
\usepackage{bm}% bold math

\newcommand{\be}{\begin{equation}}
\newcommand{\ee}{\end{equation}}
\newcommand{\beqq}{\setlength\arraycolsep{2pt}\begin{eqnarray}}
\newcommand{\eeqq}{\vspace{0cm} \end{eqnarray}}
\newcommand{\bea}{\begin{eqnarray}}
\newcommand{\eea}{\end{eqnarray}}

\newcommand{\lambdab}{\stackrel{\neg}{\lambda}}
\newcommand{\xib}{\stackrel{\neg}{\xi}}
                
\newcommand{\xb}{\bar{x}}
\newcommand{\vb}{\bar{v}}
\newcommand{\yb}{\bar{y}}
\begin{document}

\title{Some remarks on the attractor behaviour in ELKO cosmology}

\author{S. H. Pereira} \email{shpereira@gmail.com}
\author{A. Pinho S. S.} \email{alexandre.pinho510@gmail.com}
\author{J. M. Hoff da Silva}\email{hoff@feg.unesp.br}

\affiliation{Faculdade de Engenharia de Guaratinguet\'a -- Departamento de F\'isica e Qu\'imica\\ UNESP - Universidade Estadual Paulista \\ Av. Dr. Ariberto Pereira da Cunha 333 - Pedregulho\\
12516-410 -- Guaratinguet\'a, SP, Brazil}

%\maketitle

\pacs{95.35.+d, 95.36.+x, 98.80.$\pm$k, 12.60.$\pm$i}
\keywords{Dark matter, Dark energy, Cosmology, Models beyond the standard model}

%\bigskip
\begin{abstract}
Recent results on the dynamical stability of a system involving the interaction of the ELKO spinor field with standard matter in the universe have been reanalysed, and the conclusion is that such system does not exhibit isolated stable points that could alleviate the cosmic coincidence problem. When a constant parameter $\delta$ related to the potential of the ELKO field is introduced in the system however, stable fixed points are found for some specific types of interaction between the ELKO field and matter. Although the parameter $\delta$ is related to an unknown potential, in order to satisfy the stability conditions and also that the fixed points are real, the range of the constant parameter $\delta$ can be constrained for the present time and the coincidence problem can be alleviated for some specific interactions. Such restriction on the ELKO potential opens possibility to apply the ELKO field as a candidate to dark energy in the universe, and so explain the present phase of acceleration of the universe through the decay of the ELKO field into matter.
\end{abstract}

\maketitle

%%%%%%%%%%%%%%%%%%%%%%%%%%%%%%%%%%%%%%%%%%%%%%%%%%%%%%%%%%%%%%%%%%%%%%%%%%
\section{Introduction}
%ok
The relatively recent discovery of the accelerated expansion of the universe has been one of the most active research in cosmology \cite{SN}. The search for a candidate that can explain the observational data is a challenge that has drawn the attention of many researchers. In general such mysterious component is named Dark Energy (DE) (see \cite{reviewDE} for a review). The simplest candidate of DE is the cosmological constant $\Lambda$, which might explain most of the current astronomical observations. Another open question in cosmology concerns the Dark Matter (DM) problem (see \cite{reviewDM,bookDM} for a review), which is responsible for the great structures in the universe. The so called $\Lambda CDM$ model, where $CDM$ stands for Cold Dark Matter, is the best model for the present cosmology. Recent results from the Planck satellite \cite{planck} fit quite well with this model. However, from the theoretical point of view such model is plagued with some fundamental problems, thereby stimulating the search for alternative dark energy models \cite{models}. Among such alternative models, scalar dynamical fields has been proposed recently as possible candidates \cite{scalar}.
%ok

Another interesting models deal with the possibility of the coupling between DE and DM. The interaction between these completely different fluids has some important consequences, as addressing the coincidence problem, for instance. The coincidence problem could be alleviated on these models by assuming that the DE decays into DM, thus diminishing the difference between the densities of the two components through the evolution of the Universe. In a series of recent papers the possibility of a coupling between DM and DE has been considered \cite{das,Feng,DM,DE1,DE2,DE3,DE4,abwa,pavon1,wm,Jesus06,guo,Ioav,saulofernando}.
%ok

Even more recently, a special kind of non standard spinor field has also been studied both as a DM candidate (from the point of view of quantum field theory) as well as DE (in cosmological applications). This spinor field is the so-called ELKO \cite{AHL1,AHL2,AHL3}, which has some interesting and unusual properties. To begin with, this spinor field is formed by a complete set of eigenspinors of the charge conjugator operator, rendering it neutral under $U(1)$ interactions. Moreover, the field obeys only the Klein-Gordon equation. In other words it has mass dimension one. The conjugation of these characteristics made the field quite attractive from many perspectives within the cosmological setup \cite{BOE1,BOE2,BOE5,FABBRI,BOE3,BOE4,BOE6,GREDAT,BASAK1,BOE7,js,WEI,sadja,basak}. 

%ok

The possible interaction between the matter in the universe with the ELKO field have been studied from the point of view of dynamical systems \cite{WEI,sadja,basak}, and stable points of the system have been analysed from different aspects, depending on the choice of the dynamical variables. In \cite{WEI} and \cite{sadja} the stability analysis for some specific  potentials and interactions leads to attractor points just for critical points where, or the universe is totally ELKO dominated or is totally DM dominated, thus these stable points are not scaling solutions, which means they do not allow the coexistence of DM and ELKO field, which could alleviate the cosmic coincidence problem. In \cite{basak} a new choice of variables independent of the potential leads to a new set of stable points, but yet not scaling solutions. It is important to emphasise, however, that the dynamical system analysis performed in \cite{basak} starts from dynamical equations containing a subtle (but crucial) mistake. The authors of \cite{basak} analyse two different cases, and both are plagued with some misleading\footnote{See the last two paragraphs of the next section.}, which motivated us to the present work. In fact, starting from the proper equations we were able to show that there is not a stable fixed point for the underling dynamical system in the Case II of \cite{basak} while the Case I is very strict or even ill-defined. In order to circumvent this situation, we make use of an additional supposition, introducing a constant parameter related to the potential and the constraint it imposes, extracting physically relevant information about the system. 

This paper is organised as follows: Section II is somewhat a short review about the use of the ELKO field in cosmology, making contact with ref. \cite{basak}. In order to make explicit our claim about the crucial difference concerning the dynamical equations and their implications, we present in the Appendix the right (slightly modified in comparing with \cite{basak}) dynamical equations. Two different stability analysis are performed in the Sections III and IV, where the last one can alleviate the coincidence problem. In the final Section we conclude.

\section{The ELKO field in cosmology: Dynamical equations}
%ok
The ELKO spinor action in the curved spacetime is given by
\be
S={1\over 2}\int\sqrt{-g}\bigg({1\over 2}g^{\mu\nu}\big(\nabla_\mu \lambdab_{E} \nabla_\nu\lambda_{E} + \nabla_\nu \lambdab_{E} \nabla_\mu\lambda_{E} \big) - V(\lambdab_{E} \lambda_{E})\bigg)d^4x\,,\label{action}
\ee
where $V(\lambdab_{E} \lambda_{E})$ is the potential and $g\equiv$ det$g_{\mu\nu}$. The covariant derivatives acting on the ELKO spinors are $\nabla_\mu \lambdab_{E} = \partial_\mu \lambdab_{E} + \lambdab_{E} \Gamma_\mu$ and $\nabla_\mu \lambda_{E} = \partial_\mu \lambda_{E} -\Gamma_\mu \lambda_{E}$, where $\Gamma_\mu$ are the spin connections. 
%ok
The metric in a spatially flat, homogeneous and isotropic Friedmann-Robertson-Walker in a expanding universe is given by
\be
ds^2=dt^2 - a^2(t)(dx^2+dy^2+dz^2)\,.\label{ds}
\ee
%ok
The ELKO Lagrangian density can be writing as
\be
\mathcal{L}=\sqrt{-g}\bigg[{1\over 2}g^{\mu\nu}(\nabla_\mu \lambdab_{E} \nabla_\nu \lambda_{E}) -V(\lambdab_{E} \lambda_{E})\bigg]\,,\label{lagrangian}
\ee
and the equations of motion follows from a principle of least action for $\mathcal{L}$.

As has been done in recent works \cite{BOE3,BOE4,BOE6,BOE7,WEI}, we restrict the ELKO spinor field to the form $\lambda_{E}\equiv \phi(t)\xi$ and $\lambdab_{E}\equiv \phi(t)\xib$, where $\xi$ and $\xib$ are constant spinors. In \cite{js} it has been presented exact solutions to ELKO spinor in spatially flat Friedmann-Robertson-Walker expanding space times, and it has been shown that such factorisation of the time component of the ELKO field is possible for some types of scale factors.

Due to the homogeneity of the field ($\partial_i \phi =0$), the equation of motion that follows from (\ref{lagrangian}) is substantially simplified to,
\be
\ddot{\phi}+3H\dot{\phi}-{3\over 4}H^2\phi+V_{,\phi}=0\,,\label{eqphi}
\ee
where $H=\dot{a}/a$ and $V_{,\phi}\equiv dV/d\phi$. The pressure and energy density of spinor dark energy are, according to \cite{BOE6}, respectively given by
\be
p_\phi={1\over 2}\dot{\phi}^2-V(\phi)-{3\over 8}H^2\phi^2-{1\over 4}\dot{H}\phi^2-{1\over 2}H\phi\dot{\phi}\,,\label{press}
\ee
\be
\rho_\phi={1\over 2}\dot{\phi}^2+V(\phi)+{3\over 8}H^2\phi^2\, .\label{rho}
\ee

It is supposed that the universe is filled with only two components, namely a matter energy density $\rho_m$ representing the DM a and a ELKO energy density $\rho_\phi$, which could represent the DE for the late time acceleration or the inflaton field for the inflationary epoch. The Friedmann equations in a flat background, the ELKO pressure and energy density can be recast in the form\footnote{Such a decomposition on the pressure and energy density was introduced by Basak et al. \cite{basak}.}
\be
H^2 = \frac{\kappa^2}{3}\left(\rho_m +\rho_{\phi}\right)\,,\label{H2}
\ee
\be
\dot{H}=-{\kappa^2 \over 2}(\rho_m +p_m +\rho_{\phi}+p_\phi)\,,\label{Hdot} 
\ee
\be
p_\phi = X - \tilde{V}\,,\hspace{1cm}\rho_\phi = X + \tilde{V}\,,\label{rhop}
\ee
where $\kappa^2 \equiv 8\pi G$ and
\be
X={1\over 2}\dot{\phi}^2-{1\over 8}\dot{H}\phi^2-{1\over 4}H\phi\dot{\phi}\,,\label{X}
\ee
\be
\tilde{V} =  V(\phi)+{1\over 8}\dot{H}\phi^2+{1\over 4}H\phi\dot{\phi}+{3\over 8}H^2\phi^2\,.\label{Vt}
\ee%ok
Despite both new variables do not contain pure kinetic and potential elements we shall call, for simplicity, $X$ and $\tilde{V}$ as the kinetic and potential energy of the field $\phi$, respectively. The continuity equations for matter and scalar field are, respectively
\beqq
\dot{\rho}_m+3H(\rho_m + p_m)=Q\,,\label{rhom}
\eeqq
\beqq
\dot{\rho}_\phi+3H(\rho_\phi + p_\phi)=-Q\,,\label{rhophi}
\eeqq
where $Q$ stands for a possible interaction term between the DM and the ELKO field. If $Q=0$ there is no interaction and the two components evolve separately. If $Q>0$ there is the decay of ELKO field into DM, an interesting scenery at the inflation, and if $Q<0$ we have DM decaying into ELKO field (or DE), an interesting approach to late time acceleration.  The matter part is described by a perfect fluid with equation of state $p_m=(\gamma - 1)\rho_m$. 

Following \cite{basak}, it is defined the new variables
\be
x={\kappa \sqrt{X}\over \sqrt{3}H}\,, \hspace{1cm} y={\kappa \sqrt{\tilde{V}}\over \sqrt{3}H}\,, \hspace{1cm} v={\kappa \sqrt{\rho_m}\over \sqrt{3}H}\,,\label{xyv}
\ee
the Friedmann equation (\ref{H2}) can be written as a constraint equation
\be
x^2+y^2+v^2 = 1\,,\label{constr}
\ee
or in terms of the densities parameters, $\Omega_\phi + \Omega_m =1$, where
\be
\Omega_\phi = {\kappa^2 \rho_\phi\over 3 H^2}= x^2 + y^2\,, \hspace{1cm}\Omega_m = {\kappa^2 \rho_m\over 3 H^2}=v^2 \,.\label{Omega}
\ee
In order to satisfy observational data for a FRW flat universe, it will be imposed the additional condition $0\leq v^2 \leq 1$ and $0\leq x^2 + y^2 \leq 1$.

The equations (\ref{Hdot}), (\ref{rhom}) and (\ref{rhophi}) can be written as a dynamical system of the form (see the Appendix for a brief deduction):
\beqq
x'&=&(\epsilon - 3)x - {\lambda\over 2 H}{y^2\over x}-{Q_1\over x}\,,\label{xl}\\
v'&=&\bigg(\epsilon - {3\over 2}\gamma \bigg)v + {Q_1\over v}\,,\label{vl}\\
y'&=&\bigg(\epsilon +{\lambda\over 2 H}\bigg)y\,,\label{yl} 
\eeqq%ok
where 
\beqq
\epsilon \equiv -{\dot{H}\over H^2}=3x^2+{3\over 2}\gamma v^2\,,\label{ep}
\eeqq
and $'$ stands for the derivative with respect to $N\equiv \ln a$, such that $f'= \dot{f}/H$ for any function $f$. We reinforce the appearance of a $1/2$ factor in the second term of the right-hand side of Eq. (\ref{xl}). The following parameters are defined: $\lambda = {\dot{\tilde{V}}\over \tilde{V}}$ and $Q_1 = {\kappa^2 Q\over 6 H^3}$. $\epsilon$ is related to the decelerated parameter $q$ according to
\be
q\equiv -{\ddot{a}\over a H^2} =\epsilon-1\,, \label{q}
\ee%ok
so that the expansion is accelerated for $q<0$ (or $\epsilon<1$) and decelerated for $q>0$ (or $\epsilon >1$). Specifically, it is important to note that recent observational results from the Planck satellite measurements of the CMB temperature and lensing-potential power spectra \cite{planck} gives $q_0\simeq -0.527$ for the present deceleration parameter, with $\Omega_m\simeq 0.315$ and $\Omega_\Lambda \simeq 0.685$ ($31.5\%$ of dust matter in the universe and $68.5\%$ of dark energy, or cosmological constant, responsible for the present accelerated expansion). Another interesting scenery concerns the inflation, which must have $q\to -1$ (or $\epsilon \to 0$), so that the expansion could be nearly exponential or a de Sitter evolution.

%ok
The above three dynamical equations (\ref{xl}), (\ref{vl}) and (\ref{yl}) are exactly the same as obtained by Basak et al. \cite{basak}, except by the factor 2 in the denominator of the second term in the right-hand side term of Eq. (\ref{xl}). Such missing factor, as we will show later in next section, took the authors of \cite{basak} to a misplaced result about stability in this system.

It is important to notice that, indeed, the above system is not yet in a true dynamical system form, since that it contains the term $\lambda/2H$, which is clearly dependent on the dynamical variables by the term $\lambda$, which is defined as $\dot{\tilde{V}}/\tilde{V}$ and $\tilde{V}$ is explicitly $y$ dependent. There are two ways to solve this problem. First we can suppose that such term is a function of the other dynamical variables, namely $\lambda/2H=f(x,v,y)$, so we have a well-defined dynamical system. Another possibility is to setting $\lambda/2H$ as a constant, so that the dynamical system is also well-defined. According to the definition of the $\lambda$ parameter, such constant is related to the potential $V(\phi)$ of the ELKO field.

In order to study the stability of the above system, a trivial way to
satisfy $y'=0$ is take $y=0$ (which corresponds to Case I of \cite{basak}). However this condition is very restrictive, since it implies $\tilde{V}=0$, which represents a very particular choice for the potential. The case $y \neq 0$ is much
more general, which justifies our new stability analysis.

\section{Stability analysis with ${\lambda\over 2H}=-\epsilon$}

In order to turn the above system of equation in a true dynamical system and study its stability for different types of interaction term $Q$, we impose the condition\footnote{This condition corresponds to the Case II analysed by Basak et al. \cite{basak}. But, again, in \cite{basak} there is an important missing factor.} $\epsilon = -{\lambda\over 2H}$. It is easy to see that ${\lambda\over 2H}$ is defined as a dynamical quantity by means of the parameter $\epsilon$. This automatically satisfies $y'=0$, and the system (\ref{xl}) and (\ref{vl}) turns to:
\beqq
x'&=&-3xv^2 + {3\over 2}\gamma{v^2\over x}(1 -v^2)- {Q_1\over x}\,, \label{s1}\\
v' &=& 3vx^2 - {3\over 2}\gamma v(1-v^2)+{Q_1\over v}\,.\label{s2}
\eeqq%ok
The associated linearised matrix, ensured by the topological equivalence settled by the Hartmann-Grobman theorem \cite{booksystem}, is given by 
\begin{eqnarray}
\left(\begin{array}{c}
\delta x'\\ \delta v'
\end{array}\right)=M
\left(\begin{array}{c}
\delta x\\ \delta v
\end{array}\right)\,,
\label{delta_sys}
\end{eqnarray}
where
\begin{eqnarray}
M=
\left(\begin{array}{cc}
-3{v}^2 - {3\over 2}\gamma {{v}^2\over {x}^2}(1-{v}^2)+{Q_1\over {x}^2}-{1\over {x}}{\partial Q_1\over \partial {x}} & \;\; -6{v}{x} +{3\gamma {v} \over {x}}(1-2{v}^2) -{1\over {x}}{\partial Q_1\over \partial {v}}\\
 6{x}{v} + {1\over {v}}{\partial Q_1\over \partial {x}}& 3{x}^2 -{3\over 2}\gamma (1-3{v}^2)-{Q_1\over {v}^2} + {1\over {v}}{\partial Q_1\over \partial{v}}
\end{array}\right).
\label{matrix_sys}
\end{eqnarray}
$\delta x$ and $\delta y$ are the infinitesimal displacements about the fixed points. 

The stability of the system at a fixed point can be obtained from the standard analysis of the determinant ($\Delta$) and the trace ($\tau$) of the matrix $M$. According to the usual dynamical system theory, if $\Delta <0$ the eigenvalues are real and have opposite signs, hence the corresponding fixed point is a saddle point. On the other hand, if $\Delta > 0$ and $\tau <0$ the fixed point is stable, whilst if $\Delta > 0$ and $\tau > 0$ the fixed point is unstable \cite{booksystem}. The fixed points $(\bar{x},\bar{v})$ for which the above system satisfies $x'=0$ and $v'=0$ depends on the choice of the interaction term $Q$, and several possibilities will be treated in the sequel. Here we consider only the case where the matter part is pressureless, thus we take $\gamma = 1$ from now on. 

\subsection{$Q_1=0$}
%ok
Here we have $Q=0$, and consequently, there is no interaction between the standard matter and the ELKO field. Such interaction was treated by Wei \cite{WEI} with choice of the variables other than (\ref{xyv}), and no stable point was found. The fixed points of the system (\ref{s1})-(\ref{s2}) are given by $[\xb=x,\,\vb=0]$ and $[\xb=\pm \sqrt{{1\over 2}(1-\vb^2)},\,\vb=v]$. For the first fixed point we have $\Delta = 0$, so we do not have any information about the stability of the system. Furthermore, $\vb=0$ infers that $\Omega_m = 0$ and $\Omega_\phi =1$ (from (\ref{constr}) and (\ref{Omega})) a fully Dark Spinor dominated universe. The second fixed point also has $\Delta = 0$.

\subsection{$Q_1=\beta$}
%ok
In this case $\beta$ is constant. If we redefine $\beta = {3\over 2}\beta'$, we have $Q=6\beta \kappa^2 H^3 = 3\beta' H (\rho_\phi + \rho_m)$, an interaction term also treated by Wei \cite{WEI}. The fixed points are $[\xb=\pm\sqrt{{1\over 2}(1-\vb^2)-{\beta\over 3\vb^2}},\, \vb = v]$ and it is easy to show that $\Delta =0$.

\subsection{$Q_1=\beta v^2$}
%ok
In this case we have $Q=2\beta H \rho_m$. The fixed points are $[\xb=x,\,\vb=0]$ and $[\xb=x,\,\vb=\pm\sqrt{(1-2\xb^2)-{2\over 3}\beta}]$ and nothing can be said about the stability of the fixed points, since $\Delta =0$ in both cases.

\subsection{$Q_1=\beta x^2$}
%ok
At this time we have an interaction of the form $Q=\beta H (\rho_\phi+p_\phi)$. The fixed points are $[\xb=\pm\sqrt{{3\over 2}{(1-\vb^2)\over (3\vb^2+\beta)}}\vb,\, \vb = v]$, and again $\Delta =0$.

\subsection{$Q_1=\beta v x^2$}
%ok
For this kind of interaction $Q={1\over \sqrt{3}}\beta \kappa \sqrt{\rho_m}(\rho_\phi+p_\phi)$. The fixed points are $[\xb=\pm\sqrt{{3\over 2}\vb{(1-\vb^2)\over (3\vb+\beta)}},\,\vb=v]$, and it can be easily obtained that $\Delta =0$.

\subsection{$Q_1=\beta x v^2$}
%ok
In this case we have $Q=\sqrt{2\over 3}\beta \kappa \rho_m\sqrt{\rho_\phi+p_\phi}$. The fixed points are $[\xb=x,\,\vb=0]$ and $[\xb=x,\,\vb=\pm\sqrt{1-2\xb^2-{2\over 3}\beta \xb}]$ and $\Delta =0$ in both cases.

\subsection{$Q_1=\beta x^2 v^2$}
%ok
Here we have $Q={1\over 3}\beta \kappa^2 \rho_m(\rho_\phi+p_\phi)$. The fixed points are $[\xb=x,\,\vb=0]$ and $[\xb=x,\,\vb=\pm\sqrt{1-2\xb^2-{2\over 3}\beta \xb^2}]$ and, as the other cases, we have $\Delta =0$.

%ok
Having analysed all the previous cases where the determinant is always zero, we have used an algebraic manipulation software to test different functions. The functions analysed were of the types: $x^n f(v)$ and $v^n f(x)$, with $n=1,2,3,4$, and for all of them the determinant is always zero. This leads us to the conclusion that the new choice of variables keeps the same results studied by Wei \cite{WEI}, where no point of stability has been found for different types of interaction.
 
\section{Stability analysis with ${\lambda\over 2H}=-\delta$}

The above study show us that the dynamical system characterised by the
equations (\ref{xl})-(\ref{vl}) with the Friedmann constraint
(\ref{constr}) does not presents an isolated fixed point, since a null
determinant means that at least one eigenvalue is zero, and there is
either a whole line of fixed points on $x$ or $v$ axis. In order to
circumvent this situation we shall investigate the subsequent
dynamical system for the case in which another (physical) constraint
can be used to select attractor points with physical meaning. We
suppose that the potential $V(\phi)$ is such that $\lambda$ satisfies
${\lambda\over 2H}=-\delta$, where $\delta$ is a constant. 

The $\delta$ parameter just reflects our ignorance about the
potential, since it is related to potential but the specific
form of the potential is not required in this analysis. Physically,
the conditions of stability satisfied by the parameter $\delta$ will
show the ranges of possibilities for the potential in order to have a
stable system. In other words, what are the restrictions on the
potential. Besides the $\delta$ parameter, all the interactions $Q$
under analysis are characterised by a coupling constant
$\beta$. According to (\ref{rhom}) and (\ref{rhophi}), positive values
of $\beta$ correspond to positive values of $Q$, which means an
increase to  DM energy density and a decrease of the ELKO energy
density, in other words, decay of ELKO into DM particles. On the other
side, negative $\beta$ values leads to decay of DM into ELKO field.
 
The cosmic coincidence problem can be alleviated if DM and DE (here
represented by the ELKO field) could coexist for the present time of
the evolution of the universe. This implies $\rho_m\neq 0$ (which is
related to $\vb^2$) simultaneously with $\rho_\phi\neq 0$ (which is related
to $\xb^2+\yb^2$). For this
reason, in which follows,  we will be interested in fixed points
satisfying such conditions.

The corresponding dynamical system obtained from (\ref{xl})-(\ref{yl}) is:
\beqq
x'&=&3x(x^2-1+{\gamma\over 2}v^2) + {\delta\over x}(1-x^2-v^2)-{Q_1\over x}\,,\label{xll}\\
v' &=& 3vx^2 - {3\over 2}\gamma v(1-v^2)+{Q_1\over v}\,,\label{vll}
\eeqq%ok
and the fixed points are chosen such that $\bar{\epsilon}=\delta$, thus the equation for $y$ at the fixed point is $y'=0$ even for $\bar{y}\neq 0$. 

\subsection{$Q_1=0$}
For this case it is possible to find two types of fixed points, however there is only one relevant for present purpose. The first fixed point is $\bar{x}=\pm 1$, which represents $\bar{y}=0$ and $\bar{v}=0$. This case could represent only the inflationary period and is not a scaling solution. According to our previous discussion we are interested only in the case $\bar{y}\neq 0$. The other fixed point is given by $\left[\bar { x } =\frac { \sqrt { 3 }  }{ 3 } \sqrt { \delta }, \bar{y}=\sqrt { 1-\frac { 1 }{ 3 } \delta  }, \bar{v}=0 \right]$, and the conditions to guarantee stability ($\Delta>0$ and $\tau<0$) is simply $\delta<\frac{3}{2}$. As we have  $\bar{v}=0$ such fixed point is not a scaling solution too.

\subsection{$Q_1=\beta$}
By taking a constant interaction between ELKO and standard matter we find two fixed points that solve the dynamical system. But, as in the last case, there is a restriction in one of them since $\bar{y}=0$, which lead us to consider only $ \left[\bar { x } =\sqrt { \frac { 2\delta ^{ 2 }-3\delta +3\beta  }{ 6\delta -9 }  }, \bar { y } =\sqrt { \frac { -2\delta ^{ 2 }+9\delta +3\beta -9 }{ 6\delta -9 }  }, \bar { v } =\sqrt { \frac { 2\beta }{ 3-2\delta  }  } \right]$. The conditions to ensure stability are $\beta\ge-\frac{3}{8}$ and $\delta<\frac{9}{4}-\frac{1}{4}\sqrt{9+24\beta}$. However, in order to have real fixed points, namely $\bar{v}^2\ge 0\,,\;\bar{x}^2\ge 0$ and $\bar{y}^2\ge 0$, the condition turns ${3\over 4}-{1\over 4}\sqrt{9-24\beta}<\delta< {3\over 4}+{1\over 4}\sqrt{9-24\beta}$ for $0<\beta<{3\over 8}$. For $\beta \to 0$, we have $0<\delta <{3\over 2}$, and if $\beta \to {3\over 8}$ we have $\delta \to {3\over 4}$. This shows that such type of interaction can alleviate the coincidence problem if the above conditions are satisfied.

\subsection{$Q_1=\beta x^2$}
This case is similar to the last one and we have again two fixed points, being one of them also meaningless because $\bar{y}=0$. The remaining fixed point is $\left[\bar{x}=\sqrt { \frac { 3\delta-2\delta^{ 2 } }{ 9-6\delta+3\beta }  }, \bar{y}=\sqrt { \frac { 9+3\beta-6\delta-2\beta\delta+2\delta^2 }{ 9-6\delta+3\beta }  }, \bar{v}=\sqrt { \frac { 2\beta\delta }{ 9-6\delta+3\beta }  } \right]$, where it is necessary $\beta \le-\frac{3}{2}$ and $\delta < 3+\beta$ or $\beta >-\frac{3}{2}$ and $\delta < \frac{3}{2}$ for such fixed point satisfy the stability condition. However, in order to have real fixed points, these conditions reduce simply to $\beta>0$ and $0<\delta<{3\over 2}$. For the present time, where $\bar{v}^2=\Omega_m=0.315$, we have $\delta={3\over 2}{\Omega_m(3+\beta)\over \beta +3\Omega_m}$. Thus, in order to satisfy all the conditions we must have ${3\over 2}\Omega_m <\delta <{3\over 2}$ if $0<\beta <\infty$, hence scalling solutions for the present time can be obtained only if ELKO field decays into matter ($\beta >0$) and the $\delta$ parameter is limited to the above range. Under these conditions the interaction $\beta x^2$ could alleviate the coincidence problem.

\subsection{$Q_1=\beta v^2$}
The present case has a fixed available point as being $\left[\bar { x } =\frac { \sqrt { 3 }  }{ 3 } \sqrt { \delta }, \bar{y}=\sqrt { 1-\frac { 1 }{ 3 } \delta  }, \bar{v}=0 \right]$. It is easy to see from Eq. (\ref{vll}) that $\bar{v}=0$ turns the parameter $Q_1$ identically equal to zero and trivially satisfies such equation. The conditions for stability for the present fixed point are: (i) $\beta \ge -\frac{3}{2}$ if $\delta < \frac{3}{2}-\beta$; and (ii) $\beta < -\frac{3}{2}$ if $\delta < 3$. Although it has stable points it is not a scaling solution.

The another solution with $\bar{v} \neq 0$ leads to $\bar{y}=0$.

\subsection{$Q_1=\beta v^2 x^2$}
In this interaction we have three types of fixed points. One of them
with $\bar{y}=0$ and two with $\bar{y}\neq 0$. For these last two
cases we have $\left[\bar { x } =\frac { \sqrt { 3 }  }{ 3 } \sqrt {
    \delta }\,,\; \bar{y}=\sqrt { 1-\frac { 1 }{ 3 } \delta  }\,,\;
  \bar{v}=0 \right]$, under the conditions $\beta <\frac { 9-6\delta
}{ 2\delta  }$ if $0<\delta<3$ and $\beta >\frac { 9-6\delta  }{
  2\delta  }$ if $\delta<0$ for stability. Although it is a stable
point it is not a scaling solution. The last point is much more
interesting, since that $\vb \neq 0$. It is given by $\left[\bar { x }
  = \sqrt{(3-2\delta) \over 2\beta}\,,\;
  \bar{y}=\sqrt{(3-2\delta)(3+2\beta)\over 6\beta}\,,\;
  \bar{v}=\sqrt{6\delta +2\delta\beta -9\over 3\beta} \right]$. The
stability conditions leads to the following conditions: (i) $\delta
<{9\over 2(3+\beta)}$ if $\beta < -3$; (ii)
${9(\beta+6-\sqrt{-\beta^2-3\beta})\over 2\beta^2+15\beta +36}<\delta
<{3\over 2}$ or ${9\over 2(3+\beta)}<\delta <
{9(\beta+6+\sqrt{-\beta^2-3\beta})\over 2\beta^2+15\beta +36}$ if
$-{3\over 2}<\beta<0$;  and (iii) ${9\over 2(3+\beta)}<\delta <{3\over
  2}$ if $\beta > 0$. 

The above three conditions ensures the stability of the
system. However, in order to also satisfy the condition of reality of
the fixed points, namely $\vb^2\geq0$, $\xb^2\geq0$ and $\yb^2\geq0$, the only
possible condition is the last one, ${9\over 2(3+\beta)}<\delta
<{3\over 2}$ for $\beta>0$. Thus it is only possible to alleviate the
cosmic coincidence problem if $\beta$ is positive, which means the
ELKO field decaying into DM. For a small $\beta$ coupling ($\beta \to
0$), we must have $\delta \to {3\over 2}$, while for $\beta \to \infty$ we
must have $0<\delta<{3\over 2}$ in order to maintain the stability of
the system. 

For the present time for instance, where $\vb^2=\Omega_m=0.315$, we
must have
$\delta={3\over 2}{\Omega_m\beta+3\over \beta +3}$ from the fixed
point $\vb$. For $\beta \to 0$ we have $\delta \to {3\over 2}$ while
for $\beta \to \infty$ we have $\delta \to {3\over 2}\Omega_m$. Curiously, such condition is the same as the one obtained in case of the interaction $\beta x^2$ above.

We have also analysed the interactions given by $Q_1=\beta v x^2$ and $Q_1=\beta v^2 x$. For the first one it was found two stable fixed points satisfying $\bar{y}\neq 0$. For the second case there is one stable fixed point. The conditions for stability are very cumbersome, so they are omitted here.

%%%%%%%%%%%%%%%%%%%%%%%%%%%%%%%%%%%%%%%%%%%%%%%%%%%%%%%%%%%%%%%%%%%%%%%%%%
\section{Concluding remarks}

In this work it has been analysed the dynamical system concerning the study of
an interacting Dark Matter model with ELKO fields. Due to an incorrect
factor in the evolution equations present in the ref. \cite{basak}, one of
the two cases there analysed leads to an inconsistent result. In their
analysis it is possible to find out stable points for some interaction
terms between DM and the ELKO field. However, with the correct factor in
the evolution equations, we have shown that for several interaction terms there are no attractor points. 

Contrary to recent works where the potential is taken as general or assume specific forms but the systems does not present stable points or not represent scaling solutions between the ELKO field and matter, here it is assumed that the potential satisfies a differential equation characterized by a constant parameter $\delta$, and stable solutions are found. Certainly, the study of possible interaction terms between ELKO and matter fields within the scope of Friedmann-Robertson-Walker backgrounds is far from trivial. The associated dynamical system is quite involved, and extracting relevant physical information is a rather difficult task. Interestingly enough, we have found some conditions on the $\beta$ and $\delta$ parameters under which the system presents stability. Specifically, for the interactions B, C and E of the Section IV  were found fixed stable points in order to alleviate the cosmic coincidence problem. For the interactions C and E it was found that  the range of $\delta$ is related to the matter density parameter $\Omega_m$ according to ${3\over 2}\Omega_m < \delta < {3\over 2}$. Such constrain on the $\delta$ parameter, when satisfied by the potential, opens the possibility to apply the ELKO field as a candidate to dark energy in the universe, and so explain the present phase of acceleration of the universe through the decay of the ELKO field into matter.

%%%%%%%%%%%%%%%%%%%%%%%%%%%%%%%%%%%%%%%%%%%%%%%%%%%%%%%%%%%%%%%%%%%%%%%%%%

%%%%%%%%%%%%%%%%%%%%%%%%%%%%%%%%%%%%%%%%%%%%%%%%%%%%%%%%%%%%%%%%%%%%%%%%%%

\begin{acknowledgements}
SHP is grateful to CNPq - Conselho Nacional de Desenvolvimento Cient\'ifico e Tecnol\'ogico, Brazilian research agency, for the financial support, process number 477872/2010-7. JMHS thanks to CNPq (482043/2011-3; 308623/2012-6) for financial support. 
\end{acknowledgements}
%%%%%%%%%%%%%%%%%%%%%%%%%%%%%%%%%%%%%%%%%%%%%%%%%%%%%
\section*{Appendix: dynamical system equations}

%\Appendix

Here we briefly present the deduction of the dynamical system equations, namely Eqs. (\ref{xl})-(\ref{yl}). The main goal is to clarify the appearance of the missing factor 2 in equation of $x'$ from Basak~et.~al.~\cite{basak}.
 
By taking Eqs. (\ref{rhop}), the derivative $\dot{\rho_{\phi}}=\rho_{\phi}' H$ and $\rho_\phi'=X' + \tilde{V}'$ into (\ref{rhophi}) we arrive at
\be
X'+6X +\tilde{V}'=-{Q\over H}\,.\label{a1}
\ee
Taking the derivative $'$ of $x^2$ from (\ref{xyv}) we have
\be
X' = 6xx'{H^2\over \kappa^2}+6x^2{H' H\over \kappa^2}\,.\label{a2}
\ee
Using ${H'\over H}={\dot{H}\over H^2}=-\epsilon$ into (\ref{a2}) and then substituting into (\ref{a1}) it is possible, after rearranging terms, to get 
\be
x'=(\epsilon - 3)x - {\lambda\over 2 H}{y^2\over x}-{Q_1\over x}\,,\label{a3}
\ee
where $\lambda = {\dot{\tilde{V}}\over \tilde{V}}$ and $Q_1 = {\kappa^2 Q\over 6 H^3}$.

The expression for $v'$ can be derived in a similar manner. By using (\ref{rhom}) we have
\be
\rho_m' + 3 \gamma \rho_m = {Q\over H}\,.\label{a4}
\ee
Taking the derivative $'$ of $v^2$ from (\ref{xyv}) we have
\be
\rho_m' = 6vv'{H^2\over \kappa^2}+2\rho_m{H' \over H}\,.\label{a5}
\ee
As before, we write $H'$ in terms of $\epsilon$ and then substitute the result into (\ref{a4}). Doing some simple manipulations it is possible find that
\be
v'=\bigg(\epsilon - {3\over 2}\gamma \bigg)v + {Q_1\over v}\,.\label{a6}
\ee

Finally, the expression for $y'$ can be obtained by taking the derivative $'$ of $y^2$ from (\ref{xyv}) and using the above definitions for $\lambda$ and $\epsilon$, and also that $\dot{\tilde{V}}=\tilde{V}' H$. After rearrange the terms we have
\be
y'=\bigg(\epsilon +{\lambda\over 2 H}\bigg)y\,.\label{a7} 
\ee

%%%%%%%%%%%%%%%%%%%%%%%%%%%%%%%%%%%%%%%%%%%%%%%%%%%%%%%%%%%%%%%%%%%%%%%%%%

\end{document}